%% file: document.tex
\documentclass[conference]{eptcs}
\usepackage{paralist}
\usepackage[pdftex]{graphicx}

\usepackage{amsmath}
\usepackage{amsfonts} 
\usepackage{color}

\usepackage{pifont}
\newcommand{\cmark}{\ding{51}}%
\newcommand{\xmark}{\ding{55}}%
\usepackage{subfig}
\usepackage{caption}
\begin{document}

\title{Conflict Detection for Edits on Extended Feature Models using Symbolic Graph Transformation}

\author{
	Frederik Deckwerth
	\institute{ 
%		Technische Universit\"at Darmstadt\\
%		\email{geza.kulcsar@es.tu-darmstadt.de} 
	}
	\and
	G{\'e}za Kulcs{\'a}r 
	\institute{ 
%		Technische Universit\"at Darmstadt\\
%		\email{geza.kulcsar@es.tu-darmstadt.de} 
	}
	\and
	Malte Lochau
	\institute{ 
%		Technische Universit\"at Darmstadt\\
%		\email{geza.kulcsar@es.tu-darmstadt.de} 
	}
	\and
	Gergely Varr{\'o}
	\institute{ 
%		Technische Universit\"at Darmstadt\\
%		\email{geza.kulcsar@es.tu-darmstadt.de} 
	}
	\and
	Andy Sch{\"u}rr
	\institute{ 
%		Technische Universit\"at Darmstadt\\
%		\email{geza.kulcsar@es.tu-darmstadt.de} 
	}
		\institute{ 
			TU Darmstadt\\
			Real-Time Systems Lab\\
			\email{\{frederik.deckwerth|geza.kulcsar|malte.lochau|gergely.varro|andy.schuerr\}@es.tu-darmstadt.de} 
		}
}
\def\titlerunning{Conflict Detection for Edits on Extended Feature Models}
\def\authorrunning{Frederik Deckwerth, G\'eza Kulcs\'ar, Malte Lochau, Gergely Varr\'o, Andy Sch\"urr}

%\author{\IEEEauthorblockN{%
%Frederik Deckwerth\IEEEauthorrefmark{1},
%G{\'e}za Kulcs{\'a}r\IEEEauthorrefmark{1},
%Malte Lochau\IEEEauthorrefmark{1},
%Gergely Varr{\'o}\IEEEauthorrefmark{1} and
%Andy Sch{\"u}rr\IEEEauthorrefmark{1}}
%\IEEEauthorblockA{\IEEEauthorrefmark{1}
%Real-Time Systems Lab\\
%Technische Universit{\"a}t Darmstadt, Darmstadt, Germany\\
%Email: %
%\textbraceleft frederik.deckwerth,geza.kulcsar,malte.lochau,%
%gergely.varro,andy.schuerr\textbraceright@es.tu-darmstadt.de}}


\maketitle

\begin{abstract}
Feature models are used to specify variability of user-configurable systems as appearing, e.g., in software product lines. 
Software product lines are supposed to be long-living and, therefore, have to continuously evolve over time to meet ever-changing requirements.
Evolution imposes changes to feature models in terms of edit operations.
Ensuring consistency of concurrent edits requires appropriate conflict detection techniques.
However, recent approaches fail to handle crucial subtleties of extended feature models, namely constraints mixing feature-tree patterns with first-order logic formulas over non-Boolean feature attributes with potentially infinite value domains.
In this paper, we propose a novel conflict detection approach based on symbolic graph transformation to facilitate concurrent edits on extended feature models. 
We describe extended feature models formally with symbolic graphs and edit operations with symbolic graph transformation rules combining graph patterns with first-order logic formulas. 
The approach is implemented by combining eMoflon with an SMT solver, and evaluated with respect to applicability.
\end{abstract}


% \fi
%\IEEEpeerreviewmaketitle

\input{sections/sec1_introduction}
\input{sections/sec2}
\input{sections/sec3}

\input{sections/sec4}

\input{sections/sec5}

%\input{sections/sec6}
\input{sections/sec6a}
\input{sections/sec7_conclusion}

%\section*{Acknowledgment}
%%
%\begin{itemize}
%	\item TODO: CASED
%	\item TODO: IMoTEP
%	\item This work has been funded by the German Research Foundation (DFG) within the
%Collaborative Research Center (CRC) 1053 -- MAKI.
%\end{itemize}

\bibliographystyle{eptcs}
\bibliography{bib/lit}

\end{document}

%% file: sections/sec1_introduction.tex
\section{Introduction}\label{sec:intro}
In many of nowadays' application domains, 
software systems must be extensively
user-configurable in order to meet diverse
customer needs.
The inherent \emph{variability}
of those systems imposes new kinds of challenges to developers 
throughout the entire product life cycle.
Software product line engineering
constitutes a promising paradigm
to cope with the additional complexity
arising in variant-rich software systems~\cite{Czarnecki00}.
As a consequence, software product lines (SPLs) have recently
found their way from academia into real-world
application domains, such as mobile devices,
automotive and security~\cite{Weiss2008}.
An SPL comprises a family of similar,
yet well-distinguished (software) products, whose
commonality and variability are defined in terms
of \emph{features}.
Each feature, therefore, corresponds to
(1) a user-visible configuration option in the problem domain of the SPL, as well as
(2) dedicated engineering artifacts within the solution space,
composable into automatically derivable implementation variants.
Feature models are frequently used to
specify the set of relevant features of an SPL, 
together with dependencies among the features, constraining
their valid combinations within product configurations.
In particular, FODA feature diagrams are widely used during domain analysis, as they provide an 
intuitive graphical layout in terms of a tree-like
hierarchical structuring of feature nodes~\cite{Kang1990}.
Over 20 years of research has been spent to date
on developing techniques for efficiently validating crucial consistency properties
of feature models in an automated way~\cite{Benavides2010}.
Most of those approaches are limited
to feature models with Boolean feature
parameters by applying respective constraint-solving
capabilities, cf., e.g.,~\cite{Mendoncca2009,Benavides2005}. 
However, various extensions to feature models have been proposed
in order to capture all issues that are relevant for the configuration of real-world applications.
One major extension of feature models adds
complex cross-tree constraints involving non-Boolean feature attributes, e.g., 
to denote numerical configuration information
such as extra-functional properties~\cite{Passos2011}.
Until now, no generally accepted definition
of feature models extended with non-Boolean feature attribute constraints exists. 
Instead, most recent approaches 
rely on ad-hoc representations
of feature attribute constraints, e.g., being restricted
to cases that can be encoded into well-known 
constraint satisfaction problems~\cite{Benavides2005,Karatas2010,Buerdek2014}.
%%%

Designing an SPL 
for a particular application domain 
from scratch usually requires enormous domain analysis efforts and 
respective upfront investments.
Hence, an SPL is supposed to be inherently
long-living and, therefore, has to evolve
over time to prevent \emph{software aging}, imposed by ever-changing
customer needs, platform changes, new legal restrictions etc.~\cite{Parnas:1994:SA:257734.257788}.
SPL evolution, first of all,
induces \emph{edit} operations to feature model specifications
by means of changes applied to the feature diagram~\cite{McGregor2003}. 
Moreover, certain changes that occur frequently during evolution might be extracted 
as reusable change patterns~\cite{Passos2015}.
In order to ensure consistency and to prevent
structural decay, a formal framework
for feature model evolution by means of continuous
edit operation chains is required.
This becomes even more challenging
in the presence of non-Boolean feature attribute
constraints, as potential conflicts among
edit operations involving both feature-tree patterns and 
complex cross-tree constraints are hard to detect.

%%%
%
In this paper, we propose a formalization 
of feature models with complex constraints
involving non-Boolean feature attributes using
symbolic graphs as introduced in~\cite{Lazy}.
Symbolic graphs constitute a natural
extension to typed graphs 
for a concise representation of modeling languages mixing 
graphical syntax and first-order logic formulas.
This way, we are able to handle
feature diagrams with cross-tree constraints incorporating
(1) logic formulas combined with arbitrary background theories, e.g., 
linear and non-linear arithmetics, and
(2) attributes with potentially infinite value domains like
integers, real numbers, strings, etc.
As a consequence, edit operations applied to extended feature models
affecting both the feature diagram and the complex 
cross-tree constraints are 
seamlessly expressible in terms of symbolic graph transformation
rules.
Thereupon, we apply a recently proposed conflict
notion~\cite{GaM} for symbolic graph transformation rules
to ensure consistency of concurrent feature model edits.

To summarize, we make the following contributions:\begin{inparaenum}[(i)]
	\item We formalize extended feature models using symbolic graphs, naturally integrating graph patterns and logic formulas.
	\item We define edit operations on extended feature models by means of symbolic graph transformation rules.
	\item We apply a recent conflict detection notion to analyze potential conflicts among concurrent edits.
	\item We evaluate the applicability of our approach, based on an implementation combining our graph transformation tool eMoflon with the Z3 SMT solver.
\end{inparaenum}
%
%%%%%%%%%%%%%%%%%%%%%
%
%The remainder of this paper is organized as follows.
%
%\begin{itemize}
%	\item \textbf{TODO:} Background and Motivation \ref{sec:background}
%	\item \textbf{TODO:} EFM as Symbolic Graphs \ref{sec:EFMasSymbGr}
%	\item \textbf{TODO:} EFM Evolution with Symbolic Graph Transformation, Conflict Notion \ref{sec:EFMTransformation}
%	\item \textbf{TODO:} Implementation \ref{sec:impl}
%	\item \textbf{TODO:} Evaluation \ref{sec:eval}
%	\item \textbf{TODO:} Related Work \ref{sec:relatedwork}
%	\item \textbf{TODO:} Conclusion \ref{sec:conclusion}
%\end{itemize}

%% file: sections/sec2.tex
\section{Background and Motivation}\label{sec:background}

%In this section, we provide the background on extended feature models and their evolution. The concepts are illustrated on a running example from the domain of security systems.

\subsection{Feature Models}

The variability of an SPL is commonly expressed in terms of features.
%A \emph{feature} is a domain abstraction representing a distinct user visible product characteristic. The set of 
A \emph{feature} represents a distinct user-visible configuration option in the problem domain of an SPL~\cite{Czarnecki00}.
% . choosen to be part of a  distinguishable characteristic of an SPL that can be either \emph{selected} or \emph{deselected}. 
A \emph{feature model} is used to restrict the possible feature combinations by introducing logical dependencies among features.
A feature combination that fulfills these dependencies is called a \emph{valid configuration}. 
The \emph{configuration space} is the set of all valid configurations defined by a feature model.
Feature models are frequently denoted graphically as FODA feature diagrams~\cite{Kang1990}, which organize features into a hierarchical tree structure that can be enriched with additional cross-tree edges between features.
\begin{figure}[tb]
\centering
\includegraphics[width=0.48\textwidth]{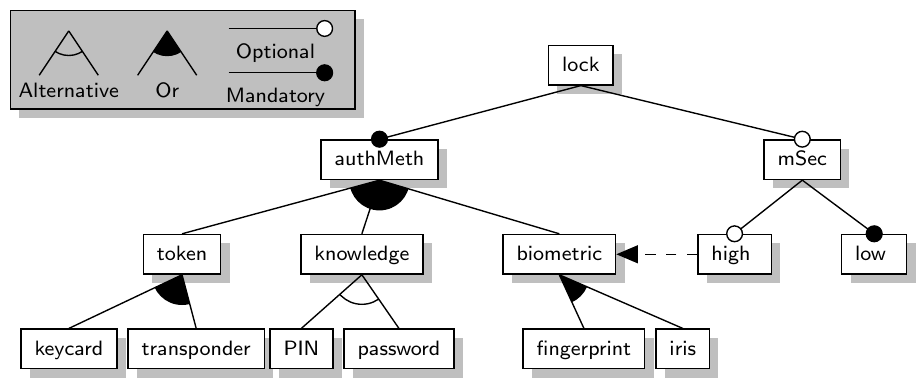} 
\caption{Feature model of an SPL for an electronic lock}
\label{fig: FODA}
\end{figure}

Figure~\ref{fig: FODA} shows a feature model representing an electronic lock in FODA notation.
%The topmost feature (\textsf{lock}) is called the root feature. % and is always part of a valid configuration. 
The feature \textsf{lock} can be equipped with different authentication devices for unlocking: \textsf{token}-based devices using a \textsf{keycard} or a \textsf{transponder}, \textsf{knowledge}-based devices for entering a \textsf{PIN} or a \textsf{password}, as well as \textsf{biometric} devices using \textsf{fingerprint} or \textsf{iris}-scan. 
The \emph{parent-child relationship} between feature nodes induces configuration constraints, i.e., the selection of a child feature requires the selection of its parent feature.
Sibling features may be arranged in \emph{groups} introducing further restrictions on valid combinations.
\begin{compactitem}
\item An \emph{alternative-group} ensures that exactly one of the child features is present whenever the parent feature is present in a configuration (e.g. either \textsf{PIN} or \textsf{password}). 

\item An \emph{or-group} claims that at least one child feature has to be present if the parent feature is present in a configuration.
In the example, an or-group is used to describe that at least a \textsf{token}-based, a \textsf{knowledge}-based or a \textsf{biometric} device has to be present in a valid lock.

\item An \emph{optional-group} consists of a single child feature that may be selected if the parent feature is selected in a configuration.
For example, the feature mission security (\textsf{mSec}) may be selected to ensure that a valid lock configuration complies with \textsf{high} or \textsf{low} security regulations.

\item A \emph{mandatory-group} also consists of one child feature and requires this feature to be present whenever its parent feature is present, e.g., a \textsf{lock} always has an authentication method (\textsf{authMeth}). 
\end{compactitem}

In addition, \emph{cross-tree edges} define dependencies between hierarchically unrelated features, e.g., in terms of binary \emph{require} and \emph{exclude} edges.
%A \emph{require} cross-tree edge from a feature $f_1$ to a feature $f_2$ requires that $f_2$ is present if $f_1$ is present. 
%An \emph{exclude} cross-tree edge between two features forbids that the features are both present in a configuration. 
For example, if the feature \textsf{high} is present in a configuration, a \textsf{biometric} authentication device must be part of the configuration, too. 
\begin{figure*}[tb]
 \centering
 \includegraphics[width=0.95\textwidth]{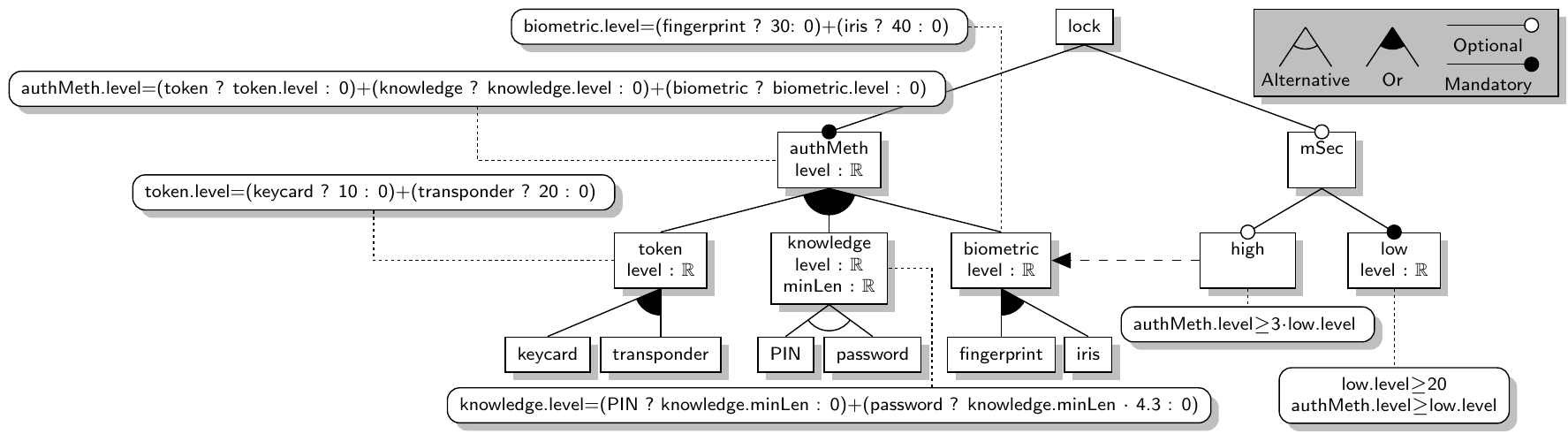}
 \caption{Extended feature model of an SPL for an electronic lock}
 \label{fig: ExFODA}
\end{figure*}

\subsection{Extended Feature Models}
In many real-world applications, it is often required to capture configuration options going beyond Boolean features of FODA feature diagrams. 
For example, it might be desirable to explicitly include the security levels of different authentication devices into the configuration process by introducing numerical measures for these characteristics. 
To handle extensions like non-Boolean configuration information, \emph{extended feature models} (EFMs) have been proposed~\cite{Czarnecki00}. EFMs extend basic FODA feature diagrams by feature attributes to augment features with additional non-Boolean configuration information. A further extension of EFMs is feature multiplicity denoted by cardinality annotations, which is out of scope here. 

A \emph{feature attribute} has a possibly infinite value domain. 
As a generalization of binary cross-tree edges, dependencies among feature attributes are denoted as complex cross-tree constraints. 
A \emph{complex cross-tree constraint} is expressed by a first-order logic formula over features and feature attributes that has to evaluate to $true$ in a valid configuration. 
A \emph{valid EFM configuration} consists of two parts: (i) a feature selection corresponding to a valid configuration and (ii) a value assignment to the attributes of the selected features, satisfying all complex cross-tree constraints relevant for those attributes.

In Figure~\ref{fig: ExFODA}, the EFM of the electronic lock is presented, extending the feature diagram from Figure~\ref{fig: FODA} with feature attributes and complex cross-tree constraints. Therein, feature attributes together with their value domains are shown in the boxes of their corresponding feature nodes. We denote complex cross-tree constraints as rounded boxes containing the first-order logic formulas. In our example, each complex cross-tree constraint is related to a feature and has to evaluate to $true$ only if its related feature is selected in a configuration. This relation may be expressed by adding further guard expressions over the related features to the constraint~\cite{Karatas2010,Benavides2005,Buerdek2014}; however, for better readability, we omit those additional clauses in our example, but rather indicate those relationships by dashed lines. 
In particular, the feature \textsf{knowledge} is extended by a feature attribute \textsf{minLen}, the value of which can be explicitly  selected by the user during configuration to adjust the minimal length required for a \textsf{PIN} or a \textsf{password}.
In addition, each feature representing an authentication method is extended with a real-valued feature attribute \textsf{level}. In contrast to \textsf{minLen}, the values of those feature attributes are not explicitly assignable by the user during configuration, but are rather implicitly derived from the selected authentication devices through complex cross-tree constraints.  
The complex cross-tree constraints in Figure~\ref{fig: ExFODA} restrict the accumulated \textsf{level} value of the selected authentication devices ($\geq 20$ if only \textsf{low} is configured, $3\ \cdot\ $\textsf{low.level} if \textsf{high} is selected). 
%This is realized by accumulating and propagating the values of the feature attributes \textsf{level} from the selected authentication methods upwards to \textsf{authMeth}, where the sum is eventually stored in \textsf{authMeth.level}. 
As feature attributes are only present in a configuration if the corresponding feature is selected, we use a \emph{default expression} to define values for the case if the corresponding feature is not selected in a configuration.  
For example, the default expression \textsf{token ? token.level : 0} evaluates to the value of \textsf{token.level} if the feature \textsf{token} is selected and it evaluates to $0$ otherwise.
By using complex cross-tree constraints, subtle dependencies among features and their attributes are expressible.
As an example, a lock with a single \textsf{keycard} device is only valid if the option mission security (\textsf{mSec}) is unselected. This is because $\mathsf{authMeth.level}$ would evaluate to 10 which is smaller than the security level 20 required by the constraints  $\mathsf{low.level} \geq 20$ and $\mathsf{authMeth.level} \geq \mathsf{low.level}$. 
As a further example, if \textsf{high} mission security is selected, the lock requires at least one further device in addition to a biometric device in order to fulfill the complex cross-tree constraint $\mathsf{authMeth.level}\geq 3 \cdot \mathsf{low.level}$.

\subsection{Concurrent Evolution of Extended Feature Models}
\label{sec: Evolution of Extended Feature Models}

As illustrated by our running example, the non-Boolean feature attributes and their corresponding cross-tree constraints add additional complexity of EFM compared to FODA diagrams, to be handled by the product-line engineer. This gets even worse by the fact that a product line is meant to be long-living, thus, requiring continuous feature model evolution~\cite{McGregor2003}.
The respective evolution steps (e.g., reflecting change requests from the customer, etc.) lead to edit operations on feature models for adding, removing or changing features, attributes and constraints~\cite{Thum2009}.

%\todo{As SPLs are supposed to be a long-term investment it is objective continuous changes during the development process. To reflect this changes the corresponding feature model needs also to be evolved. However, still our very small example gives an intention that using complex cross tree constraint can lead to very complex dependecies amongs the features. To be able 
%
%. a feature model is risky because it might impact many products and
%their users
%
%
%Still our very small example gives an intention that EFM can lead to very complex feature model specifications.
%
%
%Due to this complexity, and to reflect the \ldots a feature model is the result of an iterative process of continuous   
%
%that can be considered as a sequence of edits of the feature model. 
%Already in small sized projects the development is typically performed concurrently in a distributed setting. 
%In order to ensure consistency in an concurrent distributed setting, a formal framework for feature model evolution by means of continuous edits operation chains is required.
%In the presence of complex cross-tree constraints over feature attributes this becomes even more challenging, as potential conflicts among edits involving both the diagram patterns, as well as complex cross-tree constraints. bla bla bla catalog of typical edit operations}  

\begin{figure}
	\begin{minipage}{0.5\textwidth}
		\centering
		\includegraphics[width=\textwidth]{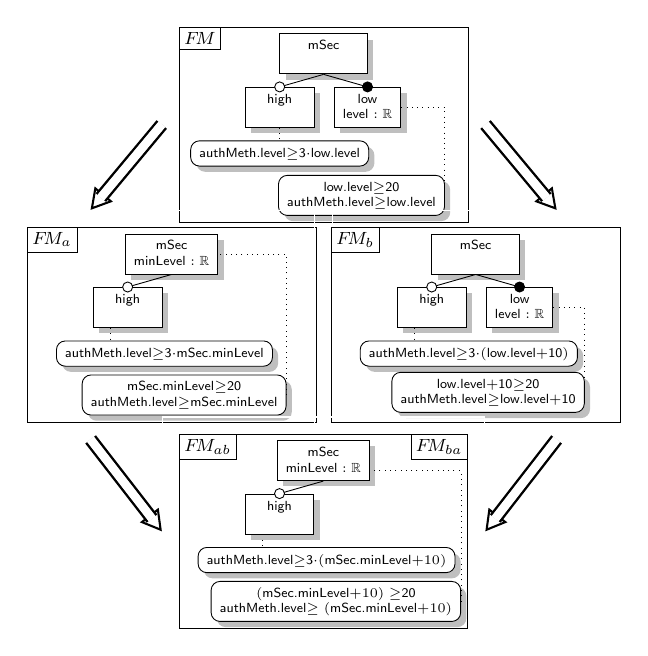} 
		\captionof{figure}{Non-conflicting evolutions of lock EFM}
		\label{fig: Com_ScaleANDMoveAttribUpDelFeature}
	\end{minipage}
	%\hspace{0.05\textwidth}
	\begin{minipage}{0.5\textwidth}
		\centering
		\includegraphics[width=\textwidth]{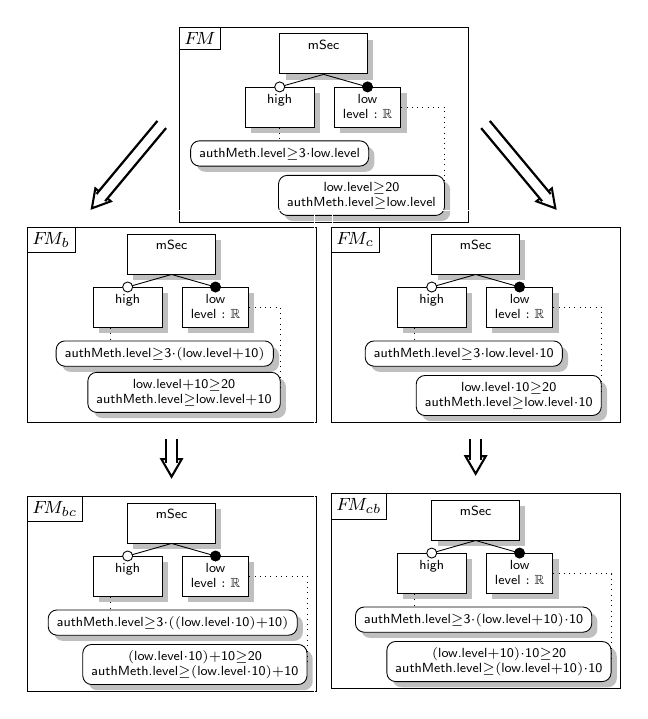} 
		\captionof{figure}{Conflicting evolutions of lock EFM}
		\label{fig: Com_ScaleANDAddLarger}
	\end{minipage}
\end{figure}

Figures~\ref{fig: Com_ScaleANDMoveAttribUpDelFeature} and~\ref{fig: Com_ScaleANDAddLarger} show examples for three different feature model edits performed on an excerpt of the EFM presented in Figure~\ref{fig: ExFODA}, referred to as $\mathit{FM}$. 
The intention of the edit (a) (shown on the left of Figure~\ref{fig: Com_ScaleANDMoveAttribUpDelFeature}) is to replace the \textsf{low} feature and its \textsf{level} feature attribute by a new feature attribute \textsf{minLevel} with the same assertions, but assigned to the feature \textsf{mSec}. 
The resulting EFM $\mathit{FM_a}$ (left of Figure~\ref{fig: Com_ScaleANDMoveAttribUpDelFeature}) is a more compact EFM, with a similar configuration semantics as $\mathit{FM}$. 
The edit (b) (right of Figure~\ref{fig: Com_ScaleANDMoveAttribUpDelFeature}) adds a constant value 10 to the value of \textsf{low.level}. The resulting EFM is $\mathit{FM_b}$ (right of Figure\ref{fig: Com_ScaleANDMoveAttribUpDelFeature}).
The aim of the edit (c) (on the right of Figure~\ref{fig: Com_ScaleANDAddLarger}) is to refactor the feature model by scaling a feature attribute value without altering the conditions imposed by the other complex cross-tree constraints. 
The result of scaling the \textsf{low.level} feature attribute by factor 10 is $\mathit{FM_c}$, shown on the right of Figure.~\ref{fig: Com_ScaleANDAddLarger}. 
% This edit can be motivated for example by new legal restrictions.     
In practice, evolution scenarios of extended feature models usually comprise many of such presumably local edits, potentially even applied concurrently by different stakeholders~\cite{Abbasi2011}. 
% \todo{SHORTEN 1-2 SENTENCES OR CAN WE LEAVE IT OUT?: The needs for performing edit operations mostly arise independently of each other and the edits should be articulated and executed with only taking their respective context and requirements into account. 
% This way of specifying edit operations results in a concurrent evolution scenario, where the identified edit operations may be parallelly applied on the feature model, i.e., the applications result in two different states of the feature model. 
% The consistency of the newly created feature model states has to be checked in order to create a new basis feature model for future evolution which comprises the effect of both parallel edits. 
% This intrinsic nature of feature model development becomes even more apparent when considering a (personally or geographically) distributed development environment for the feature model.}
It is inevitable to provide means that helps the developers to distinguish between concurrent edits whose results can be merged without obstructing feature diagram consistency, and those being potentially conflicting and require manual intervention.  
%\todo{REMOVE?: The concurrent evolution of any (even a non-extended) feature model already poses the challenges mentioned above.} 
This situation gets even more demanding in the presence of complex cross-tree constraints, as % as in our examples. 
handling concurrent EFM edits involves both feature tree patterns as well as logical reasoning on complex cross-tree constraints. For example, the edits (a) and (b) can be merged such that they result in the same EFM $\mathit{FM_{ab}}$, independent of their application order. 
This is not obvious because, after edit (a), the \textsf{low.level} attribute is removed and edit (b) is applied to the new \textsf{mSec.minLevel} attribute to obtain the same result. On the contrary, edits (b) and (c) are not arbitrarily serializable and, thus, manual prioritization might be required.

%% file: sections/sec3.tex
\section{Symbolic
Graphs and Graph Transformation for EFM Evolution}
\label{sec:FormalDescription}

In this section, we propose the use of a formal, symbolic technique to describe
extended feature models and their evolution to reason about the consistency of
concurrent edit operations performed on arbitrary extended feature models. The modeling
process consists of the following two phases: (i) Extended feature models are
described as symbolic graphs~\cite{Lazy}, which combine a graph with a first-order logic
formula, to concisely represent the feature tree structure together with the
complex cross-tree constraints (Section~\ref{sec:Modeling EFMs}). (ii) Edit operations on extended
feature models are formalized as symbolic graph transformation rules, which
provide a declarative technique to manipulate symbolic graphs (Section~\ref{sec:EFMTransformation}).

\subsection{Defining Extended Feature Models as Symbolic Graphs}
\label{sec:Modeling EFMs}

\begin{figure}[tb]
\centering
\includegraphics[width=0.48\textwidth]{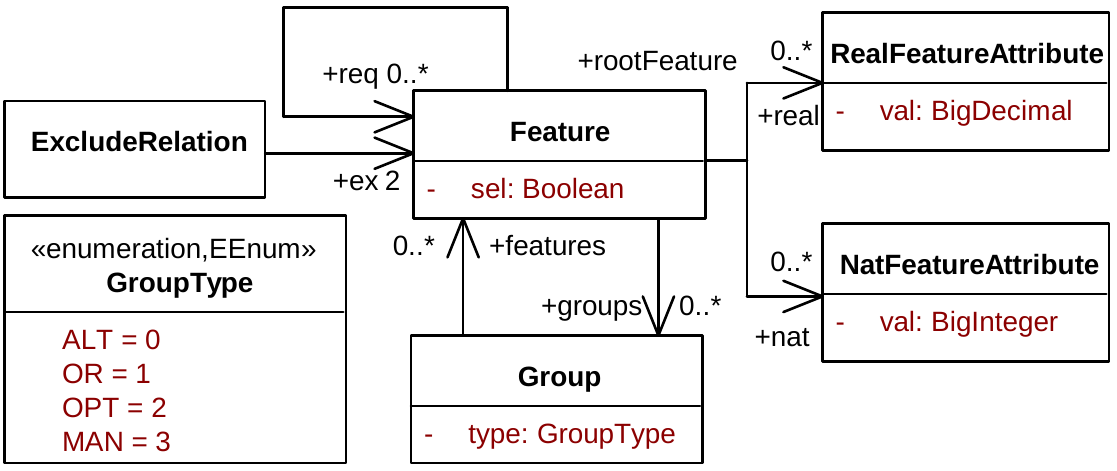}
\caption{The EFM metamodel}
\label{fig: Metamodel}
\end{figure}

To define a graph-based modeling language for EFM, we specify a metamodel for feature models with feature attributes. 
We extend this language with first-order logic for complex cross-tree constraints.
A \emph{metamodel} defines the abstract syntax for a graph-based modeling language.
% describes the abstract syntax of a modeling language and is formally considered a type graph. 
Figure~\ref{fig: Metamodel} shows the \emph{EFM metamodel} denoted as a class diagram that provides the building blocks for specifying EFMs. 
%The metamodel is denoted as UML-class diagram \cite{}. 
Nodes in the metamodel define \emph{classes} (e.g., \textsf{Feature}). 
Classes can have \emph{attributes} of a certain \emph{sort} specifying the value domain of the attribute (e.g., attribute \textsf{sel} is of sort \textsf{Boolean}).
Classes can be connected by \emph{associations} which are denoted as arrows.
Associations have \emph{cardinalities} to restrict the number of participating instances.
The EFM metamodel (Figure~\ref{fig: Metamodel}) contains the class \textsf{Feature}, which has an attribute \textsf{sel} of sort \textsf{Boolean} representing whether a feature is selected in a configuration. 
Groups are modeled by the class \textsf{Group}, which has an attribute \textsf{type} of enumeration-sort \textsf{GroupType} that defines the type of the group, i.e., alternative (\textsf{ALT}), or (\textsf{OR}), optional (\textsf{OPT}), or mandatory (\textsf{MAN}). 
The classes \textsf{RealFeatureAttribute} and \textsf{NatFeatureAttribute} for modeling feature attributes have an attribute \textsf{val} for representing the infinite domain of real and natural numbers, respectively. Further classes for feature attribute domains may be added to the metamodel if required, e.g., \textsf{StringFeatureAttribute}.
The class \textsf{Group} has association \textsf{features} to the contained child features.
A parent feature is modeled by association \textsf{groups} to the contained groups. 
The association \textsf{req} is used to define \emph{require} edges.
The \emph{exclude} edges are modeled by the class \textsf{ExcludeRelation} that has the association \textsf{ex} to the two excluding features.

A diagram specified in the modeling language defined by a metamodel is referred to as an \emph{instance model of the metamodel}. 
The nodes and edges in an instance model define \emph{objects} and \emph{links}, being instances of classes and associations of the metamodel defining their corresponding \emph{types}. 
Instances of attributes are called \emph{attribute slots} providing a location (e.g., in the memory) for a concrete \emph{attribute value} of the domain defined by the sort of the corresponding attribute in the metamodel.
By defining instance models of the EFM metamodel, we can model configurations of an EFM by creating instances of \textsf{Feature}, \textsf{Group}, \textsf{Real}- and \textsf{NatFeatureAttribute}, etc., and assigning attribute values to each attribute slot. 
However, to represent an EFM instance prior to its configuration, attribute values must be defined symbolically in terms of logic formulas, rather than by concrete value assignments. This leads to the notion of symbolic instance models.
A \emph{symbolic instance model} constitutes a natural extension of models by combining their graph-based syntax with first-order logic formulas.
% This way, we are able to naturally represent feature models with cross-tree constraints incorporating arbitrary first-order terms over attributes with potentially unbound value domains as symbolic graphs.
We denote a symbolic instance model as a pair $\langle M, \Phi_M \rangle$ consisting of an instance model $M$, whose attribute values are replaced by \emph{variables}, and a first-order logic formula $\Phi_M$ to constrain those variables.  

% a symbolic instance model can be considered as a model together with a first-order formula, where attribute values are replaced by variables that are constraint by the formula. A \emph{symbolic model} is as a tuple $\langle M, \Phi_M \rangle$ consisting of a model $M$, whose attribute values are replaced by \emph{variables} and a first-order formula $\Phi_M$, where the \emph{variable domains} are defined by the corresponding attribute type in the metamodel.  
% Instead of directly assigning values to attribute slots, a symbolic instance model contains the attribute slots are assigned to \emph{variables} that can be constraint by a first-order term included in the formula $\Phi_M$. The 
% %A symbolic model can be considered as a specification of a set of models. 
% More specifically the set of models defined by a symbolic model $\langle M, \Phi_M \rangle$ is the smallest set of models that can be obtained by substituting the variables in formula $\Phi_{FM}$ by a value from the variable domain such that $\Phi_{FM}$ evaluates to \emph{true}.

\begin{figure}
	\begin{minipage}{0.5\textwidth}
		 \centering
		 \includegraphics[width=\textwidth]{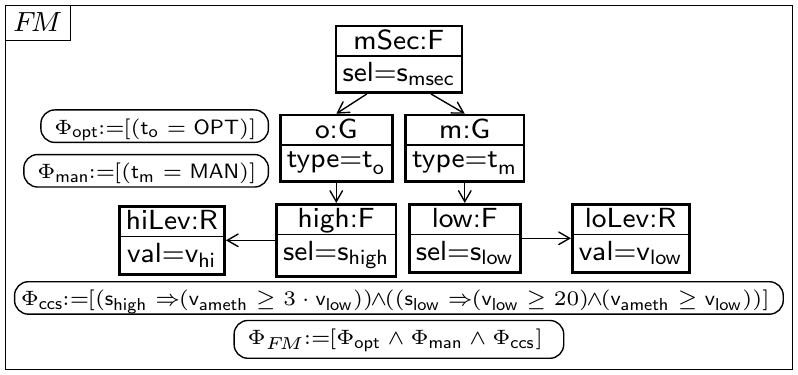}
		 \captionof{figure}{Symbolic instance model of lock EFM}
		 \label{fig: FeatureModelExcerptM}
	\end{minipage}
	%\hspace{0.05\textwidth}
	\begin{minipage}{0.5\textwidth}
			\centering
			\captionsetup[subfigure]{labelformat=empty,justification=centering}
			\subfloat[C-1]{%
				\includegraphics[width=0.3\textwidth]{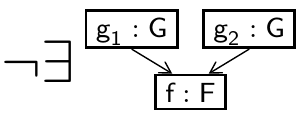}} 
			\subfloat[C-2 ]{%
				\includegraphics[width=0.3\textwidth]{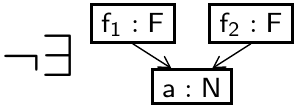}}
			\subfloat[C-3 ]{%
				\includegraphics[width=0.3\textwidth]{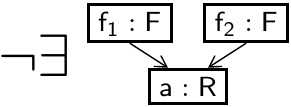}}
			\captionof{figure}{EFM well-formedness constraints}\label{fig: Well formedness constraints}
	\end{minipage}
\end{figure}

%Symbolic instance models of the EFM metamodel can concisely represent EFMs with non-Boolean feature attributes and complex cross-tree constraints.
% Moreover, we are able to obtain a precise definition of the semantic of EFMs as we can obtain any valid configuration (i.e., more specifically any instance model representing a valid configuration) from a symbolic instance model of an EFM by substituting for each variable in the formula a value such that the formula evaluates to \emph{true}.     
%which is referred to in the following as \emph{symbolic model of an EFM}.
%\begin{figure}[tb]
% \centering
% \includegraphics[width=0.45\textwidth]{figures/SymbFM0.pdf}
% \caption{Symbolic instance model of the lock EFM}
% \label{fig: FeatureModelExcerptM}
%\end{figure}
Figure~\ref{fig: FeatureModelExcerptM} shows a symbolic instance model representing an excerpt of the EFM shown in Figure~\ref{fig: ExFODA}.
The model is denoted as an object diagram where objects are represented as nodes labeled with an identifier for the object followed by a colon and the corresponding class of the object. 
Attribute slots of an object are denoted below the horizontal line and are labeled with the corresponding attribute followed by the assigned variable. 
Due to space restrictions, we abbreviate class and object names, where \textsf{F} stands for \textsf{Feature}, \textsf{G} for \textsf{Group}, \textsf{R} for \textsf{RealFeatureAttribute} and \textsf{ameth} for \textsf{authMeth}. 
The corresponding types of the links are not shown as they are unambiguously defined by the classes of the source and target objects. 
For better readability, the formula $\Phi_{\mathit{FM}}$ is partitioned in the conjunction of the terms  $\Phi_\text{opt}$, $\Phi_\text{man}$, and $\Phi_\text{ccs}$, where $\land$, $\lor$, $\Rightarrow$ and $\Leftrightarrow$ denotes the logical connectives as usual. 
The term $\Phi_\text{opt}$ sets the types of the group \textsf{o} to optional (\textsf{OPT}) and the term $\Phi_\text{man}$ sets the type of group \textsf{m} to mandatory (\textsf{MAN}).
The complex cross-tree constraints are defined in $\Phi_\text{ccs}$. 
% For example if we remove the term $s_{msec} \Rightarrow s_{low}$ from $\Phi_\text{man}$, the group \textrm{m1} is still of type \textsf{MAN}datory, but feature \textsf{mSec} can be selected without selecting feature \textsf{low}. 
To guarantee that a symbolic model of an EFM is well-formed, we define further well-formedness constraints to forbid certain patterns in a symbolic instance model. 
The \emph{EFM well-formedness constraints} are shown in Figure~\ref{fig: Well formedness constraints}:
\begin{compactenum}[{C}-1]
  \item No \textsf{Feature} is contained in two \textsf{Groups}.
  \item No \textsf{NatFeatureAttribute} is contained in two different \textsf{Features}.
  \item No \textsf{RealFeatureAttribute} is contained in two different \textsf{Features}.
\end{compactenum}

%\begin{figure} [tb]
%\centering
%\captionsetup[subfigure]{labelformat=empty,justification=centering}
%\subfloat[C-1]{%
%\includegraphics[width=0.16\textwidth]{figures/constraints/No2Groups}} 
%\subfloat[C-2 ]{%
%\includegraphics[width=0.16\textwidth]{figures/constraints/No2Nat}}
%\subfloat[C-3 ]{%
%\includegraphics[width=0.16\textwidth]{figures/constraints/No2Real}}
%\caption{EFM well-formedness constraints}\label{fig: Well formedness constraints}
%\end{figure}
%A well formedness constrains can be either a \emph{negative constraint} (C-2--C-5 and C-8--C-9) to forbid the existence of a certain pattern in a symbolic instance model

% , or a \emph{positive constraints} (C-1, C-6 and C-7) to define that all appearances of a certain pattern (denoted before the colon) can be extended to at least one of the patterns denoted after the colon.
In general, a well-formedness constraint $\neg \exists \, C$ consists of a \emph{pattern} $C$. 
A symbolic EFM instance $\langle \mathit{FM}, \Phi_{\mathit{FM}}\rangle$ satisfies a well-formedness constraint $\neg \exists\, C$ if no \emph{matching} for $C$ in $\mathit{FM}$ exists, i.e., no mapping of the elements of $C$ to the elements of a subgraph of $\mathit{FM}$ with identical graph structure and types. For instance, the symbolic EFM instance model in Figure~\ref{fig: FeatureModelExcerptM} is well-formed.

\subsection{Defining EFM Edits by Symbolic Graph Transformation}
\label{sec:EFMTransformation}

\begin{figure} [tb]
\centering
\subfloat[Edit rule $r_a$ \label{fig: r1}]{%
\includegraphics[width=0.49\textwidth]{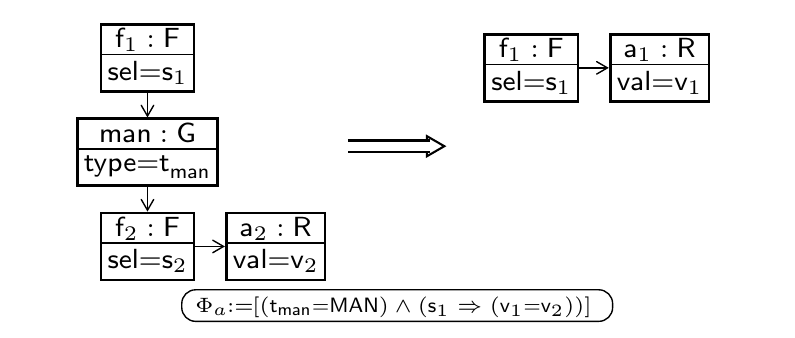}}\
\subfloat[Edit rules $r_b$ (top) and $r_c$ (bottom) \label{fig: r2}]{%
\includegraphics[width=0.49\textwidth]{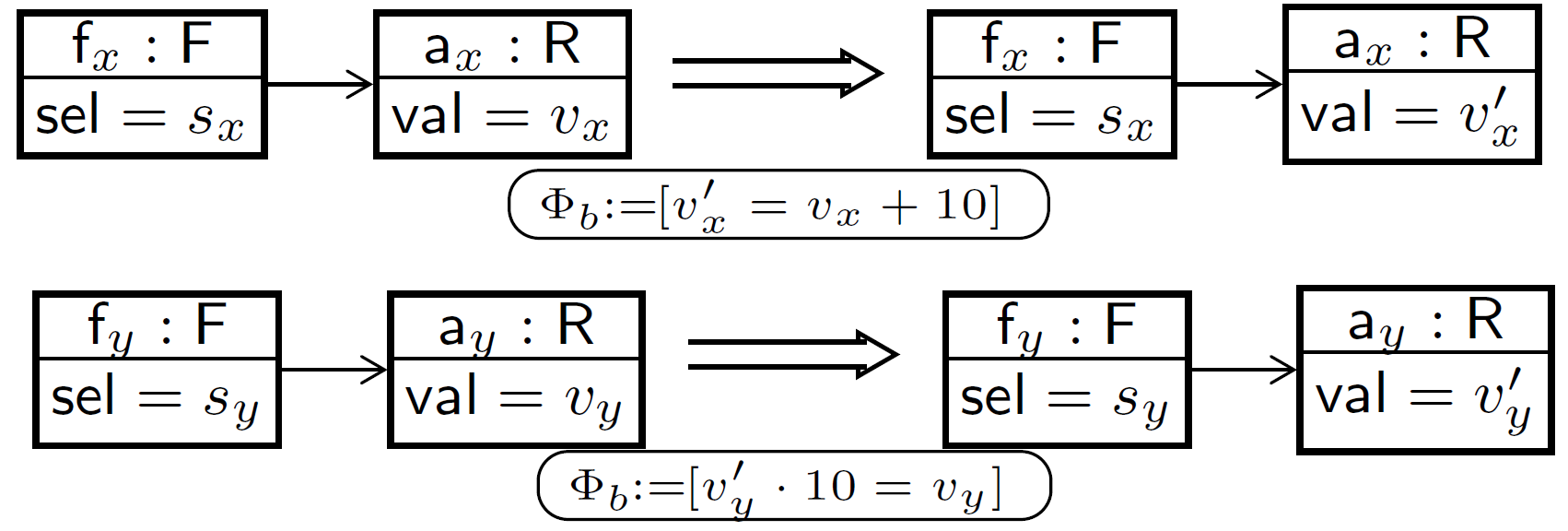}}\ 
%\subfloat[Edit rule $r_b$: \label{fig: r2}]{%
%\includegraphics[width=0.49\textwidth]{figures/Rule2}}\
%\subfloat[Edit rule $r_c$: \label{fig: r3}]{%
%\includegraphics[width=0.50\textwidth]{figures/Rule3}}\
\caption{The EFM edit rules}\label{fig: Symbolic GT-Rules for EFM edits}
\end{figure}

We define EFM edits by the rule-based technique of graph transformation.
Graph transformation (GT) provides a pattern-based manipulation of graph-based models~\cite{Fundamentals}. 
Applying a transformation rule to an instance model replaces a part of that instance model.
A \emph{symbolic GT rule} $r = (\mathit{LHS},\mathit{RHS},\Phi)$ consists of a \emph{left-hand side} pattern ($\mathit{LHS}$), a \emph{right-hand side} pattern ($\mathit{RHS}$) and a first-order logic formula $\Phi$~\cite{Lazy}. 
The $\mathit{LHS}$ of a rule defines the application context to be matched in a symbolic instance model for applying the rule.
Figure~\ref{fig: Symbolic GT-Rules for EFM edits} depicts the \emph{EFM edit rules} specifying the symbolic graph transformation rules for the EFM edits presented in Section~\ref{sec: Evolution of Extended Feature Models}. 
The $\mathit{LHS}$ pattern of rule $r_a$, depicted in Figure~\ref{fig: r1} (left to the arrow), specifies that a symbolic instance model has to contain at least one feature (to match $\mathsf{f}_1$) that is parent of a group (to match $\mathsf{man}$) that has a child feature (to match $\mathsf{f}_2$) with a \textsf{RealFeatureAttribute} (to match $\mathsf{a}_2$).
% By applying a rule the $LHS$ is replaced by the $RHS$ of the rule. 
% For example the $RHS$ of rule $r_1$ contains only the feature $\mathsf{f}_1$ and a \textsf{NatFeatureAttribute} $\mathsf{a}_1$, thus the elements $\mathsf{man}$, $\mathsf{f}_2$, $\mathsf{a}_2$ are deleted as thy have no image in the $RHS$. 
% The feature attribute \textsf{NatFeatureAttribute} $\mathsf{a}_1$ is created as it is in the $RHS$ but not in the $LHS$ and feature $f_1$ is retained as it appears in the $LHS$ and the $RHS$.

\begin{figure}[tb]
\centering
\includegraphics[width=0.5\textwidth]{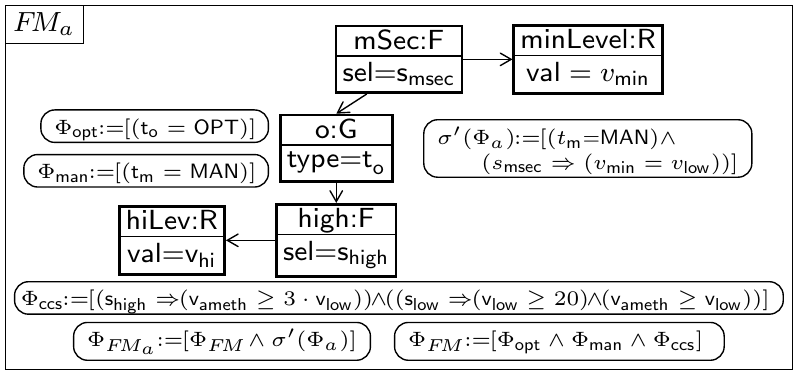}
\caption{The result $FM_a$, from applying rule $r_1$ to $FM$}
\label{fig: SymbFM1}
\end{figure}

The application of a rule $r$ at a matching of the $\mathit{LHS}$ in a symbolic instance model replaces the matching of the $\mathit{LHS}$ by the $\mathit{RHS}$ and adding the formula of the rule by conjuncting its image with the formula of the model. 
More specifically, a rule $r$ is applied at a matching $m$ to a symbolic instance model $\langle M, \Phi_M \rangle$ leading to the symbolic instance model $\langle M', \Phi_M' \rangle$ (denoted as $\langle M, \Phi_M \rangle\overset{r@m}{\Longrightarrow}\langle M', \Phi_M' \rangle$) by%
\begin{compactenum}[(i)]
\item removing elements from $M$ that can be mapped to $\mathit{LHS}$ but not to $\mathit{RHS}$,
\item adding elements that can be mapped to $\mathit{RHS}$ but not to $\mathit{LHS}$ resulting in the model $M'$, and 
\item constructing the resulting formula $\Phi_M'$ as the conjunction $\Phi_M \land \sigma'(\Phi)$ of the formula $\Phi_M$ and the image $\sigma'(\Phi)$ of the formula $\Phi$ that is obtained by substituting the variables in $\Phi$ according to the mapping of the variables from $RHS$ to $M'$.
\end{compactenum}
If the resulting formula $\Phi_M'$ is satisfiable and the resulting model $M'$ does not contain \emph{dangling links} (i.e., links whose source or target object was deleted), the rule application returns the symbolic instance model $\langle M',\Phi_M'\rangle$ and the rule application is invalid otherwise. Figure~\ref{fig: SymbFM1} shows the result of applying edit rule $r_a$ (Figure~\ref{fig: r1}) to the EFM $\mathit{FM}$ (Figure~\ref{fig: FeatureModelExcerptM}). The rule $r_a$ is applied at the matching given by the following object mapping: $(\mathsf{f}_1 \rightarrow \mathsf{mSec})$, $(\mathsf{man} \rightarrow \mathsf{m})$, $(\mathsf{f}_2 \rightarrow \mathsf{low})$, and $(\mathsf{a}_2 \rightarrow \mathsf{loLev})$. The rule is applied by%
\begin{compactenum}[(i)]
	\item removing objects $\mathsf{m}$, $\mathsf{low}$, and $\mathsf{lowLev}$ and their links and attribute slots with no image in $\mathit{RHS}$,
	\item creating a new \textsf{RealFeatureAttribute} $\textsf{minLevel}$ and assigning the new variable $v_\mathsf{min}$ to \textsf{minLevel.val},
	\item obtaining new formula $\Phi_{FM_a}$ as conjunction of $\Phi_{FM}$ and $\sigma'(\Phi_{a})$.
\end{compactenum}
The formula $\sigma'(\Phi_{a})$ is obtained by substituting the variables in the formula $\Phi_a$ of the rule according to the following mapping: $(s_1 \rightarrow s_\mathsf{msec})$, $(t_\mathsf{man} \rightarrow t_\mathsf{m})$, $(s_\mathsf{2} \rightarrow s_\mathsf{low})$, and $(v_\mathsf{2} \rightarrow v_\mathsf{low})$.
The rule application is valid as there are no dangling links and the resulting formula $\Phi_{\mathit{FM_a}}$ is satisfiable.
Note that although we delete the objects with their attribute slots, variables and their corresponding formulas are not deleted. Attribute values are changed by assigning a new variable to an attribute slot, whose value is defined by adding a new first-order clause. Based on the representation of EFM edits as symbolic graph transformation rules, we can apply a recently proposed criterion~\cite{GaM} to detect conflicting pairs of edit operations.
Two (edit) rules $r_1$ and $r_2$ are \emph{non-conflicting} if 
\begin{compactenum}[(i)]
	\item for all symbolic models $M$ and all matchings $m_1$ and $m_2$, such that rules $r_1$ and $r_2$ can be applied to $M$, the rule applications lead to the symbolic models $M_1$ and $M_2$, respectively, and
	\item there exist matchings $m_1'$ and $m_2'$ for applying $r_1$ to $M_2$ and $r_2$ to $M_1$, such that the resulting  sequences 
	$M \overset{r_1@m_1}{\Longrightarrow} M_1 \overset{r_2@m_2'}{\Longrightarrow}M_{12}$ and
	$M \overset{r_2@m_2}{\Longrightarrow} M_2 \overset{r_1@m_1'}{\Longrightarrow}M_{21}$ lead to equivalent results $M_{12}$ and $M_{21}$.   
\end{compactenum} 
Two rules are \emph{conflicting} if there exists a model and matchings with no such sequences. 

%% file: sections/sec4.tex
\section{A Conflict Detection Approach}
\label{sec:Conflict Detection Approach}

In this section, we present a conflict detection approach that operates on rules
at specification time by performing an analysis on all pairs of rules and
categorizing them as conflicting or non-conflicting. The analysis is carried out as follows: (i) For each pair of
rules, (symbolic) instance models (i.e., minimal application
contexts~\cite{ConflOfAG}) are
constructed by gluing together the left-hand sides of the rules along all
their possible type-conforming subgraphs. These instance models are considered
as representatives for a possible conflict (Section~\ref{sec:Minimal Instance Models}). (ii) The pair 
of rules under investigation is applied in both possible orders on each instance
model representative (minimal application context), and the equivalence of the 
results of these rule application sequences is checked. If any of these checks
fails (i.e., reveals the non-equivalence of the resulting instance models),
then the analyzed pair of rules is conflicting (Section~\ref{sec:Applying Rules to Minimal Contexts}).

\subsection{Constructing Minimal Instance Models}
\label{sec:Minimal Instance Models}

\begin{figure} [tb]
\centering
\captionsetup[subfigure]{justification=centering}
\subfloat[$AC^a$\label{fig: minimal contexts/gluing2}]{%
\includegraphics[width=0.16\textwidth]{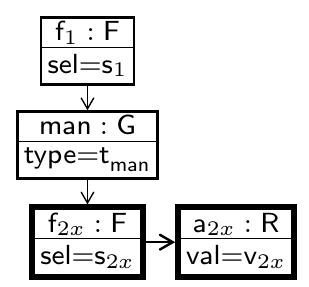}}
\subfloat[$AC^b$\label{fig: minimal contexts/gluing3}]{%
\includegraphics[width=0.16\textwidth]{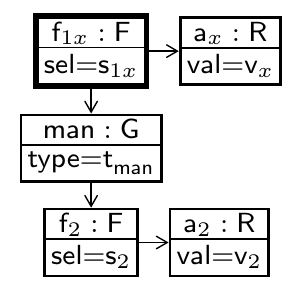}} 
\subfloat[$AC^c$\label{fig: minimal contexts/gluing5}]{%
\includegraphics[width=0.16\textwidth]{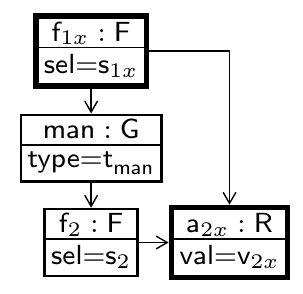}}
\subfloat[$AC^d$\label{fig: minimal contexts/gluing1}]{%
\includegraphics[width=0.24\textwidth]{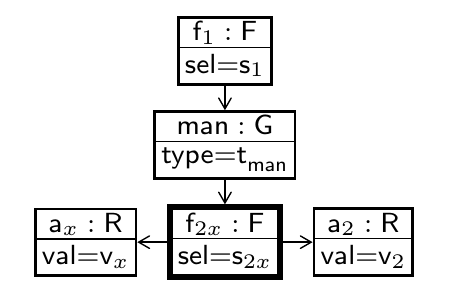}}
\subfloat[$AC^e$\label{fig: minimal contexts/gluing4}]{%
\includegraphics[width=0.24\textwidth]{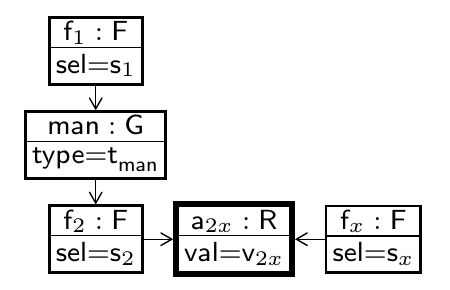}}
\caption{Minimal application contexts of the rule $r_a$ and $r_b$}\label{fig: minimal contexts}
\end{figure}

For each pair of rules, we construct instance models as minimal application
contexts by gluing together the left-hand sides of the rules along all their
possible type-conforming subgraphs. Formally, a \emph{minimal application context
$(\mathit{AC},m_1,m_2)$ of rules $r_1= (\mathit{LHS}_1,\mathit{RHS}_1,\Phi_1)$ and $r_2=
(\mathit{LHS}_2,\mathit{RHS}_2,\Phi_2)$} is a minimal instance model $\mathit{AC}$ together with the
matchings $m_1$ and $m_2$ of the left-hand sides $\mathit{LHS}_1$ and $\mathit{LHS}_2$ to the
minimal application context $\mathit{AC}$, respectively. The construction process might produce minimal instance models that
violate the EFM well-formedness constraints. These invalid instance models are
filtered out and not considered anymore in the further analysis.
Figure~\ref{fig: minimal contexts} shows all minimal application contexts for
the rules $r_a$ and $r_b$ constructed as possible gluings (shown by the bold
framed elements) of the left-hand sides $\mathit{LHS}_a$ and $\mathit{LHS}_b$. For example, the
minimal instance model $\mathit{AC}^{a}$ (Figure~\ref{fig: minimal contexts/gluing2}) is built by gluing the nodes $\mathsf{f}_2$ and $\mathsf{a}_2$ (and the edge between them) in the
left-hand side $\mathit{LHS}_a$ to the nodes $\mathsf{f}_x$ and $\mathsf{a}_x$ (and to the corresponding
edge) of the left-hand side $\mathit{LHS}_b$, respectively. The remaining part (i.e., $\mathsf{f}_1$ and $\mathsf{man}$) of that minimal instance model $\mathit{AC}^a$ originates from the
left-hand side $\mathit{LHS}_a$ and these elements do not have corresponding elements in $\mathit{LHS}_b$. 
The minimal instance model $\mathit{AC}^e$ is not a well-formed EFM as features $\mathsf{f}_2$ and
$\mathsf{f}_x$ share the same feature attribute $\mathsf{a}_{2x}$, thus, violating the constraint C-3.

\subsection{Applying Rules on Minimal Contexts}
\label{sec:Applying Rules to Minimal Contexts}

For each minimal application context $(\langle AC,true\rangle ,m_1,m_2)$ of the rules $r_1$ and $r_2$, we have to check whether we can derive application sequences
\begin{align*}
\langle AC,true\rangle \overset{r_1@m_1}{\Longrightarrow} \langle AC_1, \Phi_1\rangle \overset{r_2@m_2'}{\Longrightarrow}\langle AC_{12},\Phi_{12}\rangle \text{ and }
\langle AC,true\rangle \overset{r_2@m_2}{\Longrightarrow} \langle AC_2,\Phi_{2}\rangle \overset{r_1@m_1'}{\Longrightarrow}\langle AC_{21},\Phi_{21}\rangle
\end{align*}
such that the resulting symbolic instance models $\langle \mathit{AC}_{12},\Phi_{12}\rangle$ and $\langle \mathit{AC}_{21},\Phi_{21}\rangle$ are equivalent.
If such a sequence exists for all minimal application contexts of the two rules, the rules are non-conflicting.

\textbf{Deriving application sequences.}
First, the rules $r_1$ and $r_2$ are applied at the matchings $m_1$ and $m_2$ to $\langle AC,true\rangle $. 
If at least one rule application is invalid, the presence of this minimal application context in a model cannot lead to a conflict. 
Otherwise the rule applications lead to the symbolic instance models $\langle AC_1 , \Phi_1\rangle$ and $\langle AC_2 , \Phi_2\rangle$.
In the second step, the application sequences are derived by finding matchings $m_1'$ and $m_2'$ such that $r_1$ and $r_2$ are applicable to $\langle AC_2, \Phi_2\rangle$ and $\langle AC_1, \Phi_1\rangle$, leading to the symbolic instance models $\langle AC_{21}, \Phi_{21}\rangle$ and $\langle AC_{12}, \Phi_{12}\rangle$, respectively. 
If at least one of the matchings $m_1'$ or $m_2'$ does not exist such that the rules can be applied, $r_1$ and $r_2$ are conflicting and the analysis can be stopped.

\textbf{Checking equivalence of the results.}
Two symbolic instance models are equivalent if they have isomorphic graph parts and equivalent logic formulas.  
As the mapping of the graph parts determines the mapping of the variables as well, we have to bind those variables in $\langle AC_{12},\Phi_{12}\rangle$ and $\langle AC_{21},\Phi_{21}\rangle$ that are not assigned to any attribute slot and do not originate from $\langle AC,true\rangle $. 
These \emph{auxiliary variables} can potentially be assigned to any value (i.e., the values are only constrained by the formula) and are bound in the formulas by existential quantification.
Figure~\ref{fig: ConflictDetectionProcessExample} shows this technique for rules $r_a$ and $r_b$ with minimal application context 
$(\langle AC^a,true\rangle ,m_a,m_b)$. 
The application sequences 
\begin{align*}
\langle AC^a,true\rangle \overset{r_a@m_a}{\Longrightarrow} \langle AC^a_a, \Phi_a\rangle \overset{r_b@m_b'}{\Longrightarrow}\langle AC^a_{ab},\Phi_{ab}\rangle \text{ and }
\langle AC^a,true\rangle \overset{r_b@m_b}{\Longrightarrow} \langle AC^a_b,\Phi_{b}\rangle \overset{r_a@m_a'}{\Longrightarrow}\langle AC^a_{ba},\Phi_{ba}\rangle
\end{align*} are derived.
To compare the resulting symbolic instance models, $v_1$ and $v_c'$ in $\langle AC^a_{ab},\Phi_{ab}\rangle$ and $\langle AC^a_{ba},\Phi_{ba}\rangle$, respectively, have to be bound as they are auxiliary variables, i.e., they are not assigned to an attribute slot and do not appear in $\langle AC^a,true\rangle$.
Binding $v_1$ leads to the expression $\exists v_1 : \Phi_{ab}$ being equivalent to
\begin{equation*}
({t}_\text{man}\text{=MAN})\land({s}_{1}\Rightarrow ({v}_{x}'- 10=v_{2x}))\text{, }
\end{equation*}
as there exists a value for $v_1$ only if $v_1={v}_{x}'- 10$.
Binding $v_x'$ leads to $\exists v_x' : \Phi_{ba}$ equivalent to
\begin{equation*}
	({t}_\text{man}\text{=MAN})\land({s}_{1}\Rightarrow ({v}_{1}=v_{2x}+10))\text{.}
\end{equation*}
Thereupon, we check whether $AC^c_{ab}$ and $AC^c_{ba}$ are isomorphic, which is the case (simply mapping $\mathsf{f_1}$ to $\mathsf{f_1}$ and $\mathsf{a_1}$ to $\mathsf{a_1}$). 
Based on the mapping of the objects, we can find a variable mapping $\sigma_{ab\rightarrow ba}$: 
$(s_1 \rightarrow s_1)$, $(t_{man}\rightarrow t_{man})$, $(v_{2x}\rightarrow v_{2x})$, and $(v_x'\rightarrow v_1)$. 
In order to show that the formulas are equivalent, we have to check $\sigma_{ab\rightarrow ba}(\exists v_1 : \Phi_{ab})\Leftrightarrow \exists v_x' : \Phi_{ab}$ that is 
\begin{multline*}
({t}_\text{man}\text{=MAN})\land({s}_{1}\Rightarrow ({v}_{1}- 10=v_{2x}))
\Leftrightarrow
	({t}_\text{man}\text{=MAN})\land({s}_{1}\Rightarrow ({v}_{1}=v_{2x}+10))\text{,}
\end{multline*}
which always holds.
For the  minimal application context $AC^b$, it can be shown in the same way that no conflict occurs.
$AC^d$ does not lead to a conflict as the application of rule $r_a$ is invalid as it would produce dangling links: the link between $\mathsf{f}_{2x}$ and $\mathsf{a}_x$ becomes dangling after deleting $\mathsf{f}_{2x}$ by the rule application.
The minimal application contexts $AC^c$ and $AC^e$ are not well-formed as both violate well-formedness constraint C-3, i.e. they share a feature attribute.
Consequently, rules $r_a$ and $r_b$ are not conflicting as there exists no minimal application context that causes a conflict.
% As it is not an option to enumerate all feature models, we use technique that is relies on the notion of minimal application contexts \cite{}. 
% A minimal application context for two rules is a symbolic instance model that contains only those elements necessary for both rules to be applicable. 
%Such a minimal application context for two rules is obtained as the union the left-hand sides of both rules. 
% \todo{SOMEWERE ELSE BUT WHERE: In \cite{GAM} we have shown that it is sufficient to consider all minimal application contexts to test whether two symbolic graph transformation rules are non-conflicting.
The proposed approach is based on the fact that the application of a graph transformation rule only affects the part of a model that is included in the application context of the rule.
In order to ensure that symbolic graph transformation rules applied to symbolic instance models have only local effects, we require that the application of a rule to a symbolic instance model only adds constraints concerning fresh variables. 
More specifically, a symbolic graph transformation rule $r=(\mathit{LHS},\mathit{RHS},\Phi)$ can be handled by our approach if the expression $\forall v_1,\ldots,v_n : \Phi$ is satisfiable, where $v_1,\ldots,v_n$ are those variables appearing in the $\mathit{LHS}$ part of the rule.
The domain of any fresh variable is unaffected by this restriction.
Hence, we can modify the value of any attribute by assigning a fresh variable to the corresponding attribute slot that can be constrained arbitrarily.

%% file: sections/sec5.tex
\section{Implementation and Evaluation}\label{sec:eval}

\begin{figure}
	\begin{minipage}{0.5\textwidth}
		\centering
		\includegraphics[width=\textwidth]{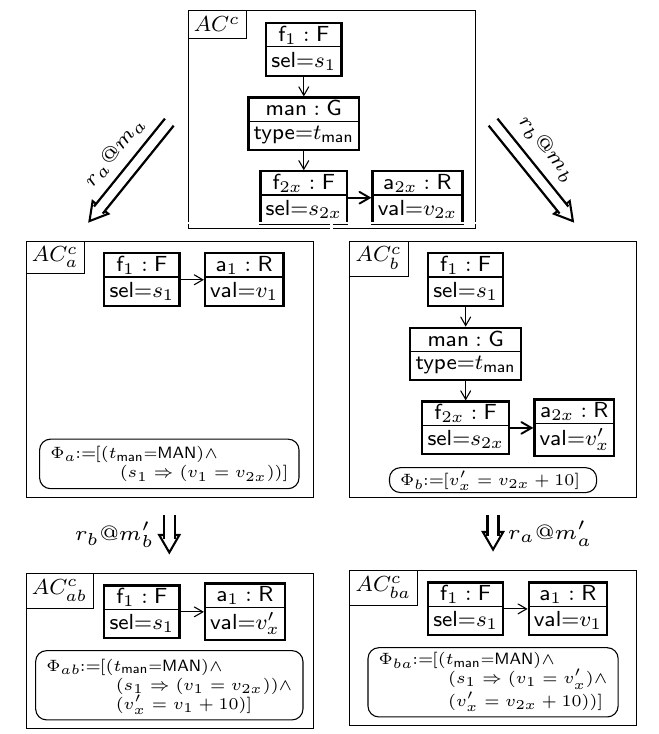}  
		\captionof{figure}{Conflict detection for rules $r_a$ and $r_b$}
		\label{fig: ConflictDetectionProcessExample}
	\end{minipage}
	\hspace{0.06\textwidth}
	\begin{minipage}{0.44\textwidth}
		\centering
		\includegraphics[width=\textwidth]{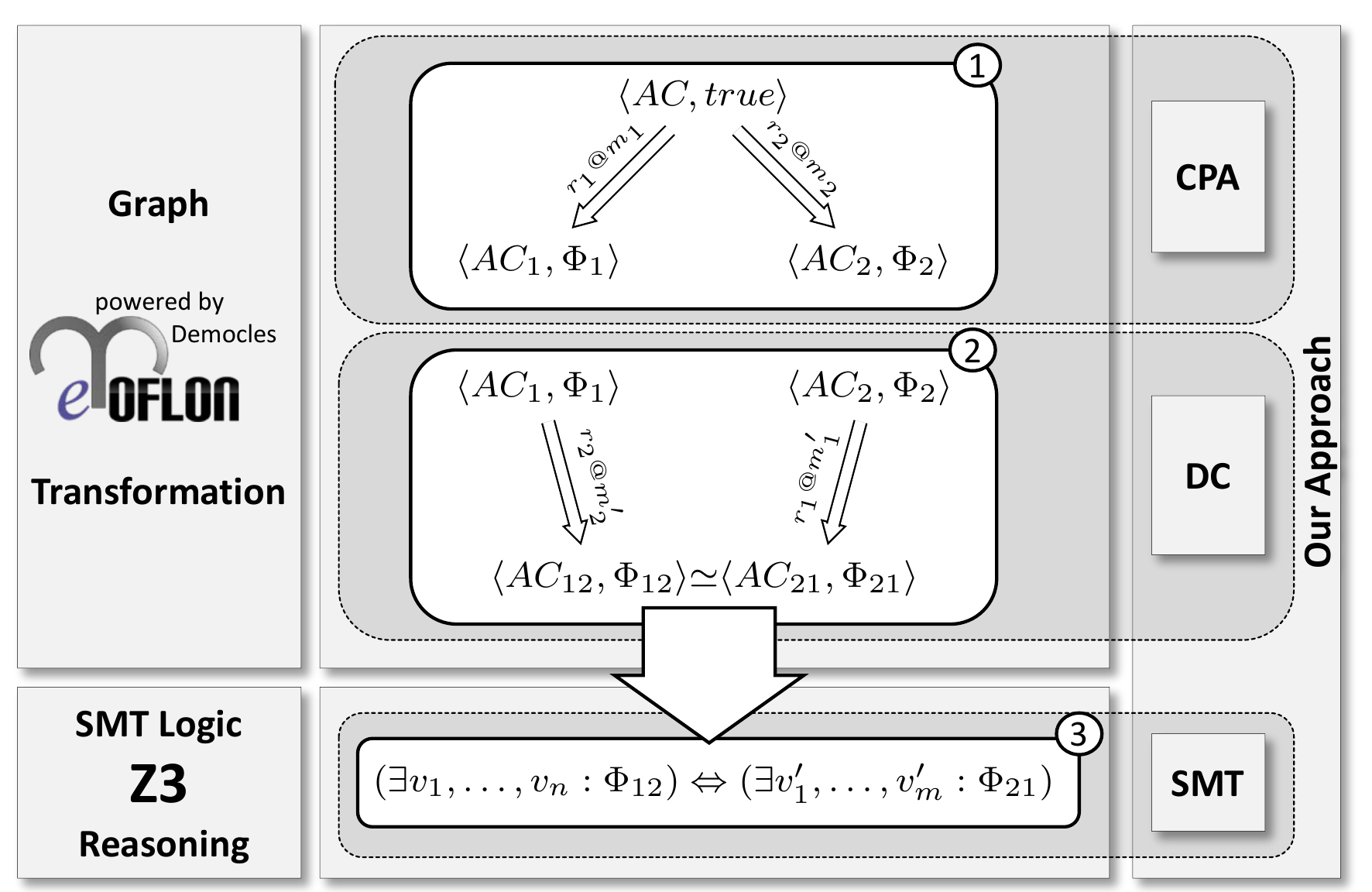}
		\captionof{figure}{Implementation architecture}
		\label{fig: Implementation}
	\end{minipage}
\end{figure}

  \begin{figure}[tb]
  	\begin{minipage}{0.5\textwidth}
  			\centering
  			\captionsetup[subfigure]{justification=centering}
  			\subfloat[CPA\label{tab:Conflict Detection Results - CPA}]{%
  				\includegraphics[height=0.22\textwidth]{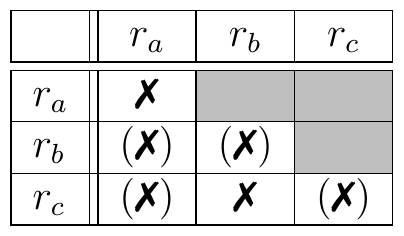}}\ \  
  			\subfloat[Our approach\label{tab:Conflict Detection Results - DC}]{%
  				\includegraphics[height=0.22\textwidth]{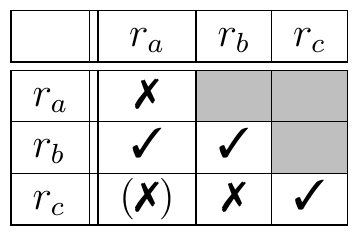}}
  			\captionof{table}{Conflict detection accuracy results}\label{tab:Conflict Detection Results}
  	\end{minipage}
  	%\hspace{0.05\textwidth}
  	\begin{minipage}{0.5\textwidth}
  			\centering
  			\includegraphics[width=\textwidth]{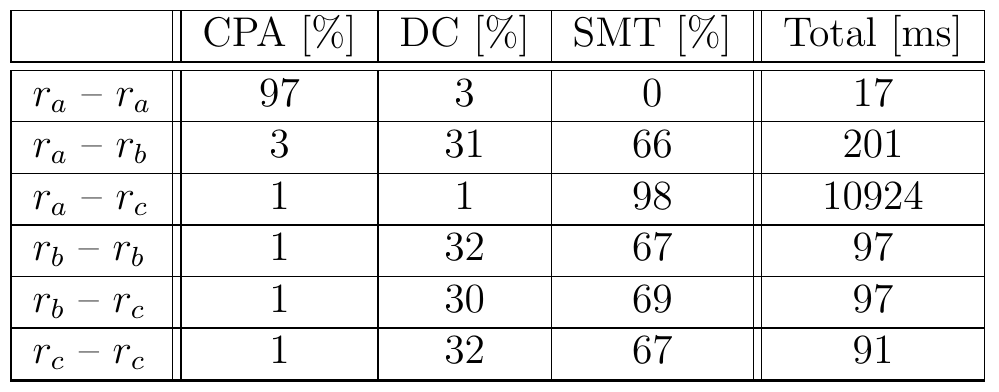}
  			\captionof{table}{Execution times of conflict detection}
  			\label{tab:mes}
  	\end{minipage}
  \end{figure}
The proposed approach is implemented by combining graph transformation with an
SMT solver. We used our model
transformation tool eMoflon~\cite{Anjorin2011} to apply rules and the Z3 SMT
solver~\cite{MoBj2008} to carry out logic reasoning. The Z3 SMT solver supports
quantification and equality for non-linear arithmetics over real
and natural numbers.
Figure~\ref{fig: Implementation} shows the architecture of our
implementation, which consists of three basic modules. The first module performs
\emph{critical pair analysis} (CPA) \ding{192}, which is a standard conflict
detection technique~\cite{Fundamentals} to derive the initial pair of rule
applications. In our approach, CPA serves as a filter to reduce the
number of minimal application contexts to be further analysed. The
direct confluence (DC) module \ding{193} searches for a second pair of rule
applications that lead to isomorphic graphs. If such a second rule application
pair is found, the variable mapping is handed over to the SMT module \ding{194}
that creates the formula and invokes the SMT solver for the equivalence check.
   We evaluate our approach by comparing it to CPA, the prevalent
   standard approach to conflict detection. In our evaluation, the accuracy (i.e.,
   the number of recognized non-conflicting rule pairs) and the execution time of
   both approaches have been assessed.
   
  Table~\ref{tab:Conflict Detection Results} shows the accuracy results for the two
  conflict detection approaches.
  A pair of non-conflicting rules is denoted by \cmark, a \xmark\ marks a
  conflicting pair of rules, and (\xmark) denotes false positives, namely,
  pairs of rules that are recognized as conflicting by the algorithms, although
  they are non-conflicting according to our definition.
  % if checked manually (such conflicts are called false positives later on). 
  The upper right of the tables has been grayed out as conflict analysis is order-insensitive.
  As shown in Table~\ref{tab:Conflict Detection Results}a, CPA categorizes
  all pairs of the edits $r_a$, $r_b$ and $r_c$ as conflicting. Note that CPA only
  considers the graph part and reports a conflict whenever one of the two analyzed edit rules 
  reassigns a variable that is in the matching of
  the other rule.
  In this way, CPA is sound (i.e., it recognizes every conflicting pair of
  rules) but less precise as it does not take the semantics of the changes
  into account.
  The accuracy results of our approach are presented in Table~\ref{tab:Conflict Detection Results}b.
  Our technique categorizes 3 pairs of rules as non-conflicting.
  Only the pair $r_a$ -- $r_c$ is incorrectly considered as a conflict.
  In this case, the SMT solver was unable to determine the equivalence of the
  formulas, i.e., the SMT solver returns \emph{UNKNOWN} (due to undecidability of first-order logics).
  Such cases are reported as conflicts (even if edit rules may
  be non-conflicting) in order to guarantee the soundness of our approach.
  % fulfills its main goal: to improve the conflict detection results of CPA by removing All except one of the false positives recoginced by CPA 
  % %3 of 4 false positives 
  % have been eliminated by our improved conflict detection.
  %, with the pair $r_a$-$r_c$ being the only remaining false positive.
  % In the case of $r_a$ -- $r_c$, the SMT solver answers with \emph{undefined}, i.e., 
  % %retain the sufficiency of our conflict condition (the pair is marked conflicting as the SMT solver was not able to decide the equivalence of the formulas, even if there is a chance that the edits are non-conflicting).
  % our approach.
  
  Table~\ref{tab:mes} presents the measured execution times for each pair
  of edits. Measurements were performed on a computer with an Intel Core
  i-7-2600-3.4GHz processor. Analysis was run 100 times for each pair of
  rules. To compensate the just-in-time optimization performed by the
  Java virtual machine, only the last 50 runs were included in the
  presented measurement results.
  The first three columns show the percentage of overall execution time for
  the three modules, and the last column contains the average of
  overall execution times in the last 50 runs.
  % single steps of the conflict detection process. 
  % The process starts with the conflict check of CPA, as a parallel independent pair is non-conflicting in any case. 
  % The second column shows the time spent with performing the transformation of the graph structure and the third one shows the time needed by the SMT solver to decide if the formulas are equivalent.
  The improved conflict detection approach is in average 4 times slower
  than CPA, where approximately $2/3$ of the additional time is used by
  the SMT solver.
  However, the overall execution times are in most cases (except for $r_a$ --
  $r_c$) below 200 ms.
  The outlier $r_a$ -- $r_c$ is caused by the fact that the SMT
  solver is unable to check the equivalence of the formulas and timeouts after 10
  seconds.
  The pair $r_a$~--~$r_a$ constitutes another extreme, where the most time is
  spent for CPA and the SMT solver is not even executed.
  The considerable time spent on CPA can be explained by the fact that
  the left-hand side of edit rule $r_a$ has the largest pattern in our setup,
  thus, gluing the pattern with itself produces the highest number of minimal application contexts.
  For $r_a$ -- $r_a$, no logic reasoning is required as accidentally, for the first minimal
  context, the derivation of the second pair
  of rule applications is not possible due to dangling links. Thus, the rules are in conflict.
  % In the case of $r_a$ -- $r_a$, the short execution time where most of the time is spent with checking CPA can be explained by the following: our improved conflict detection immediately reports that the second rule application is not feasible at all, i.e., the pair is trivially conflicting. We can also conclude from the measurement results that whenever an equivalence check is also performed by the SMT solver, this step is always the most expensive one, consuming approx. $2/3$ of the overall execution time. The outlier by $r_a$-$r_c$ is caused by the fact that the SMT solver is not able to decide equivalence in this case and timeouts after a while.
  
  \textbf{Threats to validity.} We conducted our experiments on a selected case study from the security system domain, including sample edit operations observed in a real-world application scenario. However, concerning the significance of those results with respect to other application domains, further experiments have to be conducted. Regarding the scalability of the approach, the complexity of the underlying analysis problem is mainly caused by the number and size of the rules under consideration. We assume that in a typical application scenario, the number of the rules might be much larger than in our experiments, but the size of the rules is presumably similar. Note that the approach operates on rule level, therefore, its run-time complexity is only influenced by the rule size but is independent of the size of the feature model. Concerning the overall soundness of the approach, our improved conflict detection used in this paper has been proven sound in our previous work~\cite{GaM}. Finally, threats to external validity may arise from the usage of off-the-shelf SMT solver capabilities. However, Z3 is a well-established SMT solver which is widely used in many projects and known for producing reliable results.
%  \begin{compactenum}
%  	\item Data is representative (Security example representative for structurally equivalent EFMs)
%  	\begin{compactenum}
%  		\item Example contains rules with nonlinear arithmetic on natural numbers (complex theory supported by SMT solver)
%  		\item Runntime is of high degree exponential regarding the number of elements in the rule; However operates on rule level (usually only  a couple approx. 10 elements)
%  		\item We expect the size of rules is not much larger but the number of rules increases. 	
%  	\end{compactenum}
%  	\item approach relies on SMT Solver and Graph transofrmation (both theoretically founded by Term algebras and Cathegory theory)
%  	\item conflict analysis is sound as prooven im GAM
%  	\item 
%  	
%  \end{compactenum}

%% file: sections/sec6a.tex
\section{Related Work}\label{sec:relatedwork}
%
%\nocite{*}
%
\noindent\textbf{Formalization of Extended Feature Models.}
Recent approaches for formalizing
feature model configuration 
semantics either rely on translations
into equivalent constraint problems, including
SAT~\cite{Batory2005,Mendoncca2009}, CSP~\cite{Benavides2005}, and 
BDD~\cite{Abbasi2011}, or on algebraic representations, e.g., 
using set theory~\cite{Schobbens2006}.
Extending feature models with non-Boolean
feature attributes and constraints
already has been proposed by Kang et al.
in the initial FODA feasibility study in~\cite{Kang1990} and was
further elaborated by Czarnecki et al.~\cite{Czarnecki00}. 
In Passos et al.~\cite{Passos2011}, a systematic study
on the usage of non-Boolean features in various case studies
is presented. 
However, no generally accepted syntax and 
semantics for non-Boolean features exist.

Benavides et al.~\cite{Benavides2005} propose a direct translation of feature models 
with non-Boolean
feature attributes into an equivalent CSP representation.
In contrast, B\"urdek et al.~\cite{Buerdek2014}, as well as Karatas et al.~\cite{Karatas2010} 
propose a transformation (of a restricted sublanguage) 
of non-Boolean feature constraints into Boolean feature model
fragments for applying existing 
constraint solvers~\cite{Buerdek2014,Karatas2010}.
Both approaches are limited to attributes
over finite value domains and a restricted
set of algebraic operations on attributes within constraints.

\noindent\textbf{Edits on Feature Models.}
McGregor was one of the first
who pointed out the necessity of
continuous evolution of software product lines
due to their inherently long-living nature~\cite{McGregor2003}.
Following this observation,
various researchers have proposed
approaches for systematically
evolving software product lines,
starting from changes in terms of edit operations 
to the underlying feature model.
For instance, Elsner et al. identify different types
of product-line evolution scenarios 
based on frequent changes
observed in evolving real-world 
systems~\cite{Elsner2010}.
Similar to our approach,
the EvoFM approach of Botterweck et al. 
support the specification and modularization 
of feature model edits in terms of change rules~\cite{Botterweck2010,Botterweck2014}.
In contrast, Seidl et al. define modify
patterns on feature models in terms of
model deltas and provide a mapping onto
solution space artifacts affected by the changes~\cite{Seidl2012}. 
However, both approaches are limited to predefined
collections of basic syntactic edit operations on feature models
with no support for Boolean features and respective 
constraints.

Concerning the semantic impact
of feature model changes, the approach
of Alves et al. 
ensures the preservation of feature model
configuration semantics by proposing a
catalog of sound feature model refactoring patterns~\cite{Alves2006}.
More generally, Th\"{u}m et al.
present an approach for reasoning about
the semantic impact of arbitrary feature model edits
using a SAT solver~\cite{Thum2009}.
Henard et al. present a framework 
for feature model mutation aiming at generating effective
product samples for product-line testing~\cite{Henard2013}.
The framework comprises basic mutation operators
to inject local changes into the SAT-based representation
of feature models for simulating faulty product-line changes.
Again, all those approaches are limited to
feature models with Boolean features
and corresponding constraints.
Concerning extended feature models, Quinton
et al. recently proposed an approach for
ensuring consistency-preserving evolutions
of cardinality-based feature models, again,
on the basis of a translation into SAT-based representations, whereas
non-Boolean attributes are out of scope~\cite{Quinton2014}.

\noindent\textbf{Conflict Detection on Graph Transformation Rules with Attributes.}
The approach of Cabot et al.~\cite{CabotSoSym2010} presents a fully-fledged graph transformation tool framework which also incorporates a conflict detection technique for attributed graph transformation rules. Nevertheless, the approach is based on a preceding translation of the rules into OCL expressions and, consequently, the used formalism and techniques are not suitable in our symbolic setting.
Critical Pair Analysis has been extended to attributed graph transformation on term-attributed graphs in \cite{ConflOfAG}. 
%However, in contrast to direct confluence, local confluence is undecidable even for graphs without attributes.
Although this approach can handle arbitrary attribute domains, the transformation of term-attributed graphs requires term unification to be performed at every derivation step, which restricts the practical applicability of the approach. Contrary, in the symbolic case, the formula is constructed stepwise at the syntactic level and is validated afterwards using SMT solvers. 

%% file: sections/sec7_conclusion.tex
\vspace{-.5cm}
\section{Conclusion}
\label{sec:conclusion}

In this paper, we presented a systematic approach for detecting conflicts of concurrent edit operations on extended feature models based on symbolic graph transformation to support the consistent evolution of long-living software product lines. 
The approach has been implemented by combining the graph transformation tool eMoflon with the Z3 SMT solver. Our experiments show a promising improvement concerning accuracy. We observed a remarkable reduction of false positives compared to a conventional conflict detection approach based on Critical Pair Analysis.
For future work, we plan to conduct experiments using larger rule sets gained from an industrial case study from the automation domain~\cite{Buerdek2014}. Also, we plan to investigate the possible conflicts of more than two parallel edit operations. In the unattributed case, pairwise analysis suffices, however, it is an open problem in the presence of attributes.